\documentclass[runningheads]{llncs}
\usepackage[T1]{fontenc}
\usepackage{makecell}

\usepackage{graphicx,verbatim}
\usepackage{subcaption}
\usepackage{booktabs}
\usepackage{multirow}
 \usepackage{amssymb}
 \usepackage{amsmath}
\usepackage{threeparttable}
\usepackage{hyperref}
\usepackage{color,marvosym}

\begin{document}
\title{ Learning Cross-Atlas Consistent Brain Disorder Representations via Disentangled Multi-Atlas Functional Connectivity Learning}
\titlerunning{Multi-Atlas Disentangled Learning for Brain Disease Identification}
%

\author{Anonymized Authors}  
\institute{Anonymized Affiliations \\
    \email{email@anonymized.com}}

\author{Minheng Chen \inst{1} \and Chao Cao \inst{1} \and Jing Zhang \inst{1} \and Tianming Liu \inst{2}\and Dajiang Zhu\inst{1}\textsuperscript{(\Letter)}}
%
\authorrunning{ M. Chen et al.}
\institute{ Department of Computer Science and Engineering, University of Texas at Arlington, United States\\
\email{dajiang.zhu@uta.edu}\\
\and
School of Computing,  University of Georgia, United States\\
}
\maketitle  

\begin{abstract}
Functional connectivity (FC) derived from resting-state fMRI is widely used to characterize large-scale brain network alterations in neurological and psychiatric disorders. However, FC construction critically depends on the choice of brain atlas, and different parcellations may emphasize distinct organizational features, leading to heterogeneous and sometimes inconsistent representations. Existing multi-atlas approaches partially alleviate this issue but often fuse atlas-derived features or predictions at a relatively shallow level, while single-atlas disentanglement methods do not explicitly address cross-atlas heterogeneity.
We propose Multi-Atlas Disentangled Connectivity LEarning (MADCLE), a multi-branch representation learning framework that jointly encodes FC matrices derived from different brain atlases. Rather than introducing a single explicitly shared latent variable across parcellations, MADCLE learns atlas-wise disease-related representations and encourages them to be cross-atlas consistent through distributional alignment. Meanwhile, covariate-related and atlas-dependent residual factors are modeled separately using covariate similarity supervision, atlas-specific reconstruction, and decorrelation constraints, thereby reducing the leakage of non-disease and parcellation-dependent information into the disease-related embeddings.
Experiments on the ADNI and ADHD-200 datasets suggest that MADCLE achieves competitive or improved performance compared with single-atlas baselines, multi-atlas GNN/Transformer models, and recent multi-atlas consistency frameworks. These results support the potential value of structured disentanglement for FC-based disorder identification under heterogeneous parcellation schemes. 
\end{abstract}


%
%
%
\section{Introduction}

Resting-state functional MRI (rs-fMRI) provides a non-invasive way to characterize large-scale brain functional organization by measuring spontaneous blood-oxygen-level-dependent (BOLD) fluctuations. Functional connectivity (FC), commonly computed from statistical associations between regional BOLD time series, represents the brain as an interacting network and has been widely used to study neurological and psychiatric disorders, including Alzheimer's disease (AD), attention deficit hyperactivity disorder (ADHD), autism spectrum disorder, and schizophrenia~\cite{fox2005human}. These studies suggest that disease-related alterations are often distributed across large-scale functional systems rather than confined to isolated regions, making FC a useful representation for connectome-based disorder identification.
A crucial step in FC construction is brain parcellation. Different atlases define regions of interest according to different principles, such as anatomical boundaries~\cite{desikan2006automated}, functional homogeneity~\cite{craddock2012whole}, or connectivity-informed organization~\cite{fan2016human}. Consequently, FC matrices derived from different atlases may emphasize distinct aspects of brain organization and produce complementary but heterogeneous representations~\cite{wu2025impact,xu2025multi}. While such diversity can enrich disease-related information, it also introduces atlas-induced variability: the same subject may exhibit different connectivity patterns depending on the chosen parcellation, which can affect downstream classification performance and reduce the robustness of learned representations.

Recent multi-atlas learning methods attempt to alleviate this problem by leveraging complementary information from multiple parcellations~\cite{chu2022multi,mahler2023pretraining,xu2025multi}. Many existing approaches, however, mainly integrate atlas-derived features or predictions through fusion, while providing limited explicit modeling of atlas-dependent residual variation. More recent consistency-aware methods encourage agreement across atlases, but the separation between disease-relevant information and atlas-dependent effects remains insufficiently explored. In parallel, graph disentanglement methods have been introduced to separate disease-related representations from covariates such as age, sex, and acquisition site~\cite{zhang2025graph}. Yet these methods are typically developed under a single-atlas setting and do not directly address how disease-related representations should be coordinated across heterogeneous parcellation schemes.

These observations motivate a structured multi-atlas representation learning framework that can encourage disease-related information to be consistent across atlases while separately modeling covariate-related and atlas-dependent factors. In this work, we propose Multi-Atlas Disentangled Connectivity LEarning (MADCLE) for FC-based brain disorder identification. MADCLE processes FC matrices from different parcellation schemes with atlas-specific encoders and factorizes each atlas representation into disease-related, covariate-related, and atlas-dependent components. Rather than imposing a single shared latent variable, MADCLE learns atlas-wise disease representations and encourages their cross-atlas consistency through distributional alignment. Covariate-informed supervision, atlas-specific reconstruction, and decorrelation constraints are further used to reduce the leakage of non-imaging and parcellation-dependent information into disease-related embeddings.
We evaluate MADCLE on ADHD-200 and ADNI, covering neurodevelopmental and neurodegenerative disorder classification under heterogeneous parcellation schemes. Experiments show that MADCLE achieves competitive or improved performance over single-atlas and multi-atlas baselines, supporting the value of structured multi-atlas disentanglement for FC-based disorder identification.
\begin{figure}[htb]

  \centering
  \centerline{\includegraphics[width=\linewidth]{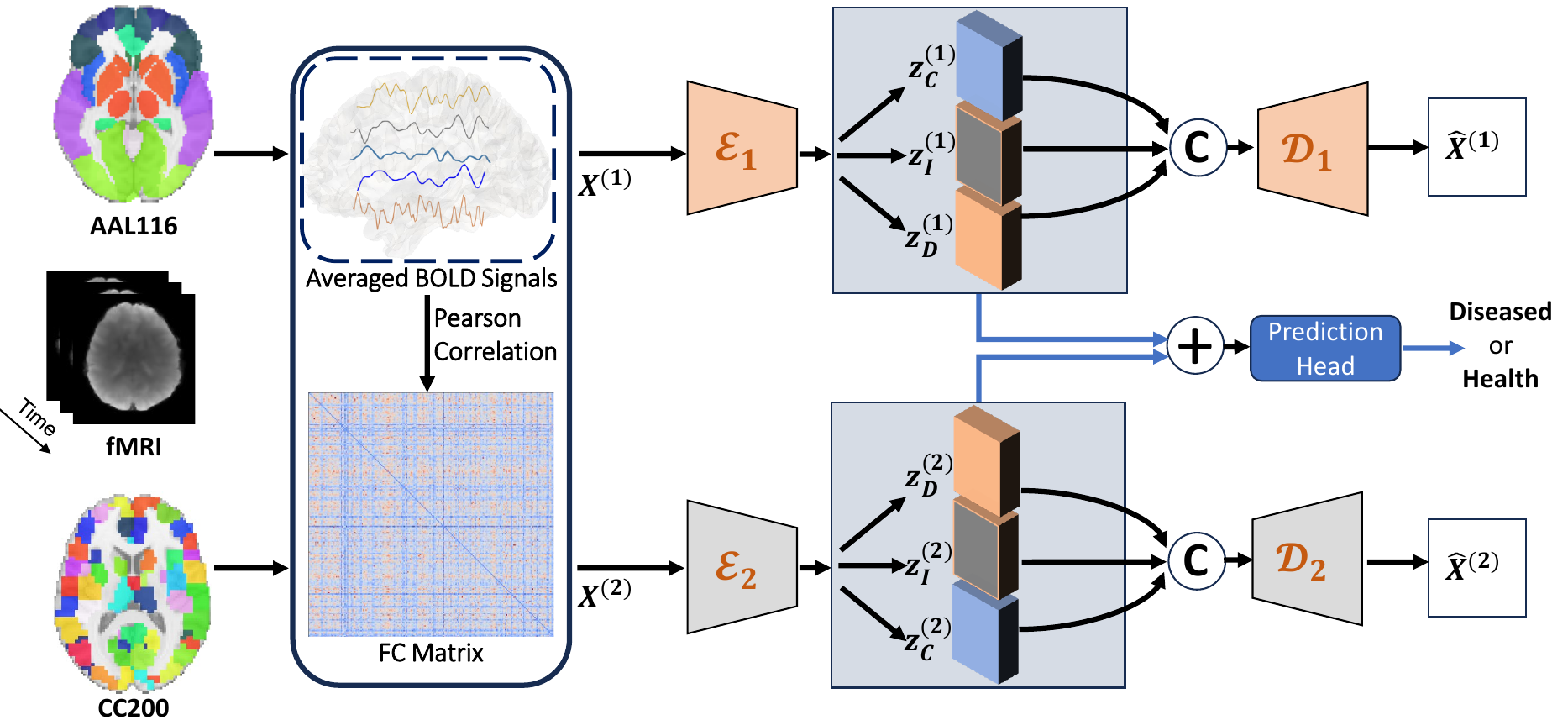}}
\caption{Overview of MADCLE. FC matrices from different parcellations are encoded by atlas-specific branches. Each branch produces disease-related, covariate-related, and atlas-dependent embeddings. Disease-related embeddings are encouraged to be cross-atlas consistent and are aggregated for diagnostic prediction, while the remaining factors regularize covariate and atlas-dependent variability. }
\label{fig:overview}
\end{figure}
\section{Methodology}
As shown in Fig.~\ref{fig:overview}, the proposed framework takes FCs derived from two parcellation schemes as input. 
The two parcellations comprise $v_1$ and $v_2$ brain regions respectively, leading to FC matrices with different spatial resolutions. After removing self-connections by setting the diagonal elements to zero and applying row-wise normalization, the resulting matrices are denoted as $X^{(k)} \in \mathbb{R}^{v_k \times v_k}, \quad k \in \{1,2\}$.
For each FC matrix, we have an independent encoder $\mathcal{E}_k(\cdot)$, where the encoder takes $X^{(k)}$ as input and outputs three disentangled latent representations:
$\left(z_D^{(k)},\, z_C^{(k)},\, z_I^{(k)}\right) = \mathcal{E}_k\!\left(X^{(k)}\right)$, where each latent vector lies in $\mathbb{R}^{d}$.
Specifically, $z_D$ encodes disease-related connectivity patterns that are discriminative for diagnostic prediction, $z_C$ captures variability associated with non-disease covariates, and $z_I$ represents atlas-specific characteristics that are independent of both disease and covariate effects.
The three latent vectors derived from the same atlas are subsequently concatenated and passed through an atlas-specific decoder $\mathcal{D}_k(\cdot)$, yielding a reconstructed FC $\hat{X}^{(k)}$.

\subsection{Network Architecture}
Each atlas-specific encoder $\mathcal{E}_k$ is implemented as a transformer-based network operating on row-wise tokenized FC matrices. Specifically, given an input FC $X^{(k)} \in \mathbb{R}^{v_k \times v_k}$ each row is treated as a node token representing the connectivity profile of a brain region. 
The input tokens are first linearly projected into a latent embedding space, \textit{i.e.}, $ h_i^{(k)} = W_{\mathrm{in}}\, x_i^{(k)}, \quad i = 1,\dots,v_k$,  and augmented with learnable node identity embeddings to preserve region-specific information. 
The node identity embedding process is defined as $\tilde{h}_i^{(k)} = h_i^{(k)} + \mathrm{MLP}\!\left(h_i^{(k)} + e_i^{(k)}\right)$, where $e_i^{(k)}\in \mathbb{R}^d$ is a  learnable token.
The resulting token embeddings are then processed by a stack of transformer encoder layers with multi-head self-attention, enabling the model to capture higher-order interactions between brain regions.
A global readout is obtained by average pooling across all node tokens, yielding a compact graph-level representation. 
This representation is subsequently fed into three parallel projection heads to generate disentangled latent embeddings corresponding to disease-related ($z_D^{(k)}$), covariate-related ($z_C^{(k)}$), and atlas-specific ($z_I^{(k)}$) factors.
To encourage the latent representations to retain sufficient information about the original FC structure, an atlas-specific decoder $\mathcal{D}_k(\cdot)$ is employed for reconstruction. The three disentangled latent vectors are concatenated and passed through a multilayer perceptron to reconstruct the FC matrix $\hat{X}^{(k)}\in \mathbb{R}^{v_k \times v_k}$. This reconstruction pathway acts as a structural regularizer, preventing degenerate solutions and promoting meaningful disentanglement across latent factors.
For disease prediction, only the disease-related representations are utilized. Specifically, the disease embeddings from the two atlas branches are aggregated via summation, $\tilde{z}_D=z_D^{(1)} +z_D^{(2)}$, and subsequently fed into a lightweight classification head composed of fully connected layers with nonlinear activation. This design encourages diagnostic prediction to rely primarily on disease-related information that is consistent across atlases, while reducing the influence of covariate-related and atlas-dependent factors.
\subsection{Multi-Atlas Disentangled Representation Learning}
A reconstruction loss is applied to the FC matrices reconstructed by the decoders
to ensure that the learned latent representations preserve information from the
original functional connectivity:
\begin{equation}
\mathcal{L}_{recon}
=
\|\hat{X}^{(1)}-X^{(1)}\|_F^2
+
\|\hat{X}^{(2)}-X^{(2)}\|_F^2 
\end{equation}
To explicitly supervise the covariate-related representation, we first min-max normalize age within the training set and apply the same normalization to validation and test subjects. Within each mini-batch, we construct a covariate similarity matrix \(S \in \mathbb{R}^{B \times B}\) based on age, sex, and acquisition site:
\begin{equation}
S_{ij} =
\frac{1}{3}
\left[
(1-|a_i-a_j|)
+
\mathbf{1}[\mathrm{sex}_i=\mathrm{sex}_j]
+
\mathbf{1}[\mathrm{site}_i=\mathrm{site}_j]
\right]
\end{equation}
where \(S_{ij} \in [0,1]\).
We then enforce the normalized cosine similarity between concatenated covariate embeddings
\(\tilde{z}_{C,i}=[z^{(1)}_{C,i};z^{(2)}_{C,i}]\) to match \(S\):
\begin{equation}
\mathcal{L}_{sim}
=
\|\hat{S}-S\|_F^2,
\quad
\hat{S}_{ij}
=
\frac{1}{2}
\left(
1+
\frac{\tilde{z}_{C,i}^{\top}\tilde{z}_{C,j}}
{\|\tilde{z}_{C,i}\|\|\tilde{z}_{C,j}\|}
\right)
\end{equation}
To encourage cross-atlas consistency in disease- and covariate-related representations, the distributions of the corresponding latent components across atlases are aligned using Jensen--Shannon (JS) divergence at the batch level.
For any $z\in \{ z_D,\, z_C\}$, we have:
\begin{equation}
    P = \frac{1}{B} \sum_{i=1}^{B} \mathrm{softmax}\!\left(z_i^{(1)}\right),
\qquad
Q = \frac{1}{B} \sum_{i=1}^{B} \mathrm{softmax}\!\left(z_i^{(2)}\right)
\end{equation}
And the alignment loss can be formulated as follow:
\begin{equation}
\mathcal{L}_{\mathrm{align}} = \mathrm{JS}\!\left(P_D \,\|\, Q_D\right) + \mathrm{JS}\!\left(P_C \,\|\, Q_C\right)
\end{equation}
Finally, to disentangle atlas-specific characteristics from disease and covariate effects, we penalize the cross-covariance between $z_I$ and $\{z_D,z_C\}$ using a Frobenius-norm regularization:
\begin{equation}
\mathcal{L}_{\mathrm{atlas}}
=
\sum_{k=1}^{2}
\left(
\left\| \mathrm{Cov}\!\left(z_I^{(k)},\, z_D^{(k)}\right) \right\|_{F}^{2}
+
\left\| \mathrm{Cov}\!\left(z_I^{(k)},\, z_C^{(k)}\right) \right\|_{F}^{2}
\right)
\end{equation} 
Together, the final objective function is:
$    \min_{\theta}\;
\mathcal{L}_{\mathrm{ce}}
    +
\lambda_r \mathcal{L}_{\mathrm{recon}}
+
\lambda_s \mathcal{L}_{\mathrm{sim}}
+
\lambda_a \mathcal{L}_{\mathrm{align}}
+ 
+\lambda_i \mathcal{L}_{\mathrm{atlas}}$, 
where $\theta$ denotes the set of model parameters, $\mathcal{L}_{\mathrm{ce}}$ is the cross-entropy loss for disease classification, and $\lambda_r$, $\lambda_s$, $\lambda_a$, and $\lambda_i$ are scalar hyperparameters that balance the contributions of the corresponding loss terms.
\section{Experiments and Results}
\subsection{Experimental Setup}
We partition each dataset into 70\% for training, 10\% for validation, and 20\% for testing. 
For evaluation on the test set, we select the epoch that achieves the highest AUROC score on the validation set.
All reported performances are the average of 10 random runs on the test set with the standard deviation.

\noindent\textbf{Datasets.}
We evaluated MADCLE on two rs-fMRI cohorts: ADHD-200 for binary ADHD diagnosis and ADNI for four-class classification among AD, early mild cognitive impairment (EMCI), late mild cognitive impairment (LMCI), and normal control (NC). 
For ADHD-200, we used rs-fMRI data processed by the Athena Pipeline from 8 imaging sites. After quality control and removal of questionable subjects, 276 ADHD subjects and 351 healthy controls were retained. 
For ADNI, 659 subjects, including 132 AD, 179 EMCI, 157 LMCI, and 191 NC subjects, were selected after quality control, with preprocessing following~\cite{zhu2014connectome}. 
For each subject, FC matrices were constructed using two complementary parcellations: the anatomical AAL116 atlas~\cite{tzourio2002automated} and the functional CC200 atlas~\cite{craddock2012whole}. Regional mean BOLD signals were extracted under both atlases, and Pearson correlation matrices were computed; only subjects with valid paired AAL116-CC200 FCs were retained. 
AAL116 and CC200 were used as the primary atlas pair due to their complementary anatomical and functional parcellation principles. The BN273 atlas~\cite{fan2016human} was additionally introduced in the ablation study to evaluate whether the proposed framework remains competitive under alternative atlas combinations.
For the ADNI four-class task, AUROC, sensitivity, and specificity were computed using macro-averaged one-vs-rest metrics, while ACC denotes overall classification accuracy.

\noindent\textbf{Baselines.}
We use 5 single-atlas methods and 5 multi-atlas methods as baselines to evaluate our proposed MADCLE, including:
(1) Conventional machine learning models: Logistic Regression (LR) and eXtreme Gradient Boosting (XGBoost). These models take the flattened lower-triangle connectivity matrix as vector input, instead of using the whole brain network, since the FC is a symmetric matrix.
(2) Deep Learning-based models for brain network analysis: BrainNetCNN~\cite{kawahara2017brainnetcnn}, BNT~\cite{kan2022brain} and CP-SSM~\cite{chen2025core}.
(3) Multi-atlas baselines: MultiRF (Random Forest using concatenated lower-triangular FC features from multiple atlases), MultiSVM (Support Vector Machine with the same concatenated features), as well as MGRL~\cite{chu2022multi}, METAformer~\cite{mahler2023pretraining}, and DMAA~\cite{han2025dual}. 
For MultiRF and MultiSVM, we adopted the multi-atlas regional feature selection strategy proposed in~\cite{min2014multi}.

\noindent\textbf{Implementation details.}
Each atlas-specific encoder $\mathcal{E}$ is implemented as a two-layer Transformer encoder,  the decoder $\mathcal{D}$ is implemented as an 2-layer MLP that takes the concatenated latent vector and outputs a vector which is reshaped into the reconstructed matrix.
Unless otherwise specified, the dimension of latent vector is set to 64, $\lambda_r$, $\lambda_s$, $\lambda_a$, and $\lambda_i$ are set to 0.5, 0.5,0.2 and 0.2 respectively.
All models are trained using AdamW with learning rate $5\times10^{-4}$ and weight decay $10^{-4}$ using a batch size of 128 and a maximum of 200 epochs.
To stabilize disentanglement, the decorrelation term is introduced with a linear warm-up over the first 20 epochs.
\subsection{Results}
\noindent\textbf{Comparison analysis.}
Table~\ref{tab:model} compares the proposed MADCLE with classical machine learning methods, single-atlas deep models, and recent multi-atlas baselines on the ADHD-200 and ADNI datasets.
On ADHD-200, MADCLE achieves an AUC of 68.9\(\pm\)2.7 and an ACC of 67.5\(\pm\)1.7, yielding competitive performance compared with both single-atlas and multi-atlas baselines. 
Compared with strong multi-atlas baselines, the improvement in AUC is modest, but the results suggest that explicitly modeling cross-atlas consistency and atlas-dependent residual information can help stabilize disease-related representations. 
On the ADNI four-class classification task, MADCLE achieves an AUC of 89.9\(\pm\)1.3 and an ACC of 72.5\(\pm\)2.9, showing improved performance over the evaluated baselines. 
Overall, these results suggest the potential usefulness of structured multi-atlas disentanglement for FC-based disorder identification across heterogeneous cohorts.
\begin{table*}[t]
\centering
\begin{threeparttable}
\caption{Performance comparison  with different baselines on ADHD200 and ADNI with the best and second-best values in \textbf{boldface} and \underline{underline}.}
\label{tab:model}
\begingroup
\setlength{\tabcolsep}{0.3pt}   
\renewcommand{\arraystretch}{1.1}
\scriptsize                  

\begin{tabular*}{\textwidth}{@{\extracolsep{\fill}}%
l l
c c c c 
c c c c 
@{}}
\toprule
\multicolumn{2}{c}{\multirow{2}{*}{\textbf{Methods}}}& 
\multicolumn{4}{c}{\textbf{ADHD200}} &
\multicolumn{4}{c}{\textbf{ADNI}}\\ 
\cmidrule(lr){3-6}\cmidrule(lr){7-10} &&
\textbf{AUC} & \textbf{ACC} & \textbf{SEN} & \textbf{SPE}  &
\textbf{AUC} & \textbf{ACC} & \textbf{SEN} & \textbf{SPE} \\
\midrule
\multirow{6}{*}{\makecell{AAL\\116}}
  & LR     & 60.3$\pm$5.8 & 59.0$\pm$4.4 & 56.9$\pm$6.3 & 60.6$\pm$7.4 & 74.9$\pm$2.9 & 47.6$\pm$4.7 & 45.7$\pm$5.2 & 82.2$\pm$1.6  \\
  & XGBoost   & 59.4$\pm$4.6 & 59.1$\pm$3.9 & 46.2$\pm$6.0 & 69.0$\pm$7.9 & 81.5$\pm$2.1 & 59.4$\pm$3.0& 58.4$\pm$3.2& 86.2$\pm$1.0  \\
  & BrainNetCNN   & 65.2$\pm$8.5 & 62.4$\pm$7.0 & 56.0$\pm$8.7 & 69.0$\pm$7.2 & 75.8$\pm$1.4 & 53.1$\pm$1.0 & 53.1$\pm$1.4 & 84.3$\pm$0.3  \\
  & BNT   & 68.0$\pm$5.3 & 62.3$\pm$5.8 & 58.6$\pm$6.6 & 62.8$\pm$5.4 & 84.3$\pm$2.7 & 64.3$\pm$3.8 & 64.0$\pm$4.0 & 87.9$\pm$1.3  \\
  & CP-SSM   & 67.3$\pm$4.7 & 61.8$\pm$4.3 & 51.4$\pm$8.2& 69.0$\pm$7.3 & 87.0$\pm$1.1 & 68.4$\pm$1.0 & 68.3$\pm$0.1 & 89.4$\pm$0.1  \\
\addlinespace[2pt]
\hline
\multirow{6}{*}{\makecell{CC\\200}}
  & LR     & 62.2$\pm$4.5 & 59.4$\pm$3.9 & \underline{60.9$\pm$4.3} & 58.2$\pm$6.9  & 77.9$\pm$2.4& 52.4$\pm$3.4& 50.8$\pm$3.8& 83.8$\pm$1.2\\
   & XGBoost     & 65.6$\pm$5.3 & 62.3$\pm$3.0 & 44.6$\pm$4.2 & 76.1$\pm$4.7  & 83.2$\pm$1.9 & 60.2$\pm$4.2 & 59.2$\pm$4.3 & 86.5$\pm$1.4 \\
     & BrainNetCNN   & 58.7$\pm$5.3 & 57.3$\pm$6.2 & 46.4$\pm$7.9 & 63.5$\pm$8.5& 81.0$\pm$1.3 & 60.2$\pm$2.5 & 59.6$\pm$2.5 & 86.7$\pm$0.9 \\
  & BNT   & 65.9$\pm$5.9 & 61.1$\pm$6.3 & 46.5$\pm$6.3 & 71.7$\pm$5.3 & 87.1$\pm$2.7 & 67.4$\pm$4.2 & 67.1$\pm$4.4 & 89.0$\pm$1.4  \\
  & CP-SSM   & 66.1$\pm$4.5 & 62.1$\pm$5.2 & 58.9$\pm$9.5 & 61.1$\pm$6.3 & \underline{89.0$\pm$2.3} & 68.8$\pm$4.3 & 68.5$\pm$4.6 & 89.5$\pm$1.4  \\
  \addlinespace[2pt]
   \hline
\multirow{6}{*}{\makecell{Multi\\Atlas}}
  & MultiRF     & 65.8$\pm$4.3 & 64.4$\pm$5.8 & 46.0$\pm$8.6& \textbf{78.6$\pm$6.0}  & 76.9$\pm$1.2 & 55.3$\pm$2.6 & 55.1$\pm$1.5& 84.8$\pm$1.3 \\
   & MultiSVM     & 67.5$\pm$3.7 & \underline{64.8$\pm$3.2} & 49.8$\pm$5.4 & 76.5$\pm$4.7 & 87.1$\pm$2.1& 65.2$\pm$2.6 & 65.2$\pm$3.1 & 88.3$\pm$1.7 \\
     & MGRL   & \underline{68.4$\pm$2.1} & 62.3$\pm$4.1 & \textbf{68.4$\pm$6.9} & 57.6$\pm$8.2 & 78.8$\pm$7.0 & 55.8$\pm$1.5 & 56.0$\pm$1.8 & 85.2$\pm$0.5  \\
    & METAFormer   & 66.1$\pm$3.6 & 64.7$\pm$6.1 & 56.7$\pm$9.6 & 70.8$\pm$9.3 & 88.5$\pm$3.2 & \underline{69.8$\pm$5.0} & \underline{69.6$\pm$5.1} & \underline{89.8$\pm$1.7}
    \\
    & DMAA   & 67.3$\pm$8.6 & 60.6$\pm$3.0 & 38.4$\pm$9.9 & \underline{77.9$\pm$9.9} & 82.4$\pm$1.2 & 60.9$\pm$2.1 & 61.0$\pm$1.9 & 86.9$\pm$0.1  \\
 \cmidrule{2-10}
    & \textbf{MADCLE}   & \textbf{68.9$\pm$2.7} & \textbf{67.5$\pm$1.7} & 54.9$\pm$9.1 & 77.2$\pm$9.9 & \textbf{89.9$\pm$1.3} & \textbf{72.5$\pm$2.9} & \textbf{72.1$\pm$3.0} & \textbf{90.7$\pm$1.0}  \\
\bottomrule
\end{tabular*}
\endgroup
\end{threeparttable}
\end{table*}

\noindent\textbf{Ablation and sensitivity studies.}
We first conduct an ablation study on the ADHD-200 dataset to assess the contribution of each component in MADCLE, as summarized in Table 2.
To further examine the sensitivity of MADCLE to atlas selection, we additionally introduced the BN273 atlas~\cite{fan2016human} and evaluated all pairwise atlas combinations. BN273 was used as an additional validation atlas rather than the primary atlas pair in the main comparison. The results suggest that MADCLE remains competitive across different atlas pairs, although the optimal atlas combination may vary.
Removing \(z_I\), \(L_{recon}\), \(L_{sim}\), or \(L_{align}\) generally reduces AUC or ACC, suggesting that these components contribute to the overall behavior of the framework. In particular, excluding $\mathcal{L}_{\mathrm{recon}}$ results in a notable drop in both AUC and accuracy, indicating that reconstruction plays a crucial role in preserving meaningful FC structure. The effect of \(L_{atlas}\) is more modest, mainly reflected in AUC and specificity rather than sensitivity, indicating that the decorrelation constraint may influence the sensitivity-specificity trade-off instead of uniformly improving all metrics.
 Excluding covariate-related constraints or cross-atlas alignment further reduces performance, demonstrating that each component contributes synergistically to effective multi-atlas disentanglement.
\begin{table}[h!]
\centering
\caption{Ablation study of MADCLE on the ADHD-200 dataset.}
\label{tab:ablation}
\renewcommand{\arraystretch}{1.3}
\setlength{\tabcolsep}{5pt}

\begin{tabular}{ccc|c|c|c|c}
\hline
\textbf{CC200} & \textbf{AAL116} & \textbf{BN273} &
\textbf{AUC} & \textbf{ACC} & \textbf{SEN} & \textbf{SPE} \\ 
\hline
\checkmark & \checkmark & & 68.9$\pm$2.7 & 67.5$\pm$1.7 & 54.9$\pm$9.1 & 77.2$\pm$9.9  \\
\checkmark & &   \checkmark& 70.1$\pm$2.5 & 69.0$\pm$1.7 & 54.2$\pm$9.2 &80.6$\pm$9.8  \\
 & \checkmark &\checkmark  & 70.1$\pm$1.4  & 68.7$\pm$0.8 & 54.9$\pm$10.8  & 79.4$\pm$8.6 \\
\hline
\multicolumn{3}{l|}{CC200 $+$ AAL116} & \textbf{68.9$\pm$2.7} & \textbf{67.5$\pm$1.7} & \underline{54.9$\pm$9.1} & \underline{77.2$\pm$9.9} \\
\multicolumn{3}{l|}{CC200 $+$ AAL116 (w/o $z_{I}$)} & 64.9$\pm$1.5 &66.2$\pm$1.5&53.8$\pm$7.9 &71.1$\pm$8.9 \\
\multicolumn{3}{l|}{CC200 $+$ AAL116 (w/o $\mathcal{L}_{recon}$)} &67.5$\pm$2.8 &65.6$\pm$1.8 &52.9$\pm$13.6 &75.5$\pm$9.3 \\
\multicolumn{3}{l|}{CC200 $+$ AAL116 (w/o $\mathcal{L}_{atlas}$)} &\underline{68.5$\pm$2.4} &\textbf{67.5$\pm$2.1} &\textbf{55.6$\pm$9.6} &76.6$\pm$7.1 \\
\multicolumn{3}{l|}{CC200 $+$ AAL116 (w/o $\mathcal{L}_{sim}$)} &67.7$\pm$1.9 &66.2$\pm$2.0 &50.7$\pm$9.6 &\textbf{78.2$\pm$8.4} \\
\multicolumn{3}{l|}{CC200 $+$ AAL116 (w/o $\mathcal{L}_{align}$)} & 66.7$\pm$2.5 &\underline{67.2$\pm$1.1}&\underline{54.9$\pm$9.9} &71.9$\pm$7.2 \\
\hline
\end{tabular}
\end{table}
\begin{figure}[h!]

  \centering
  \centerline{\includegraphics[width=\linewidth]{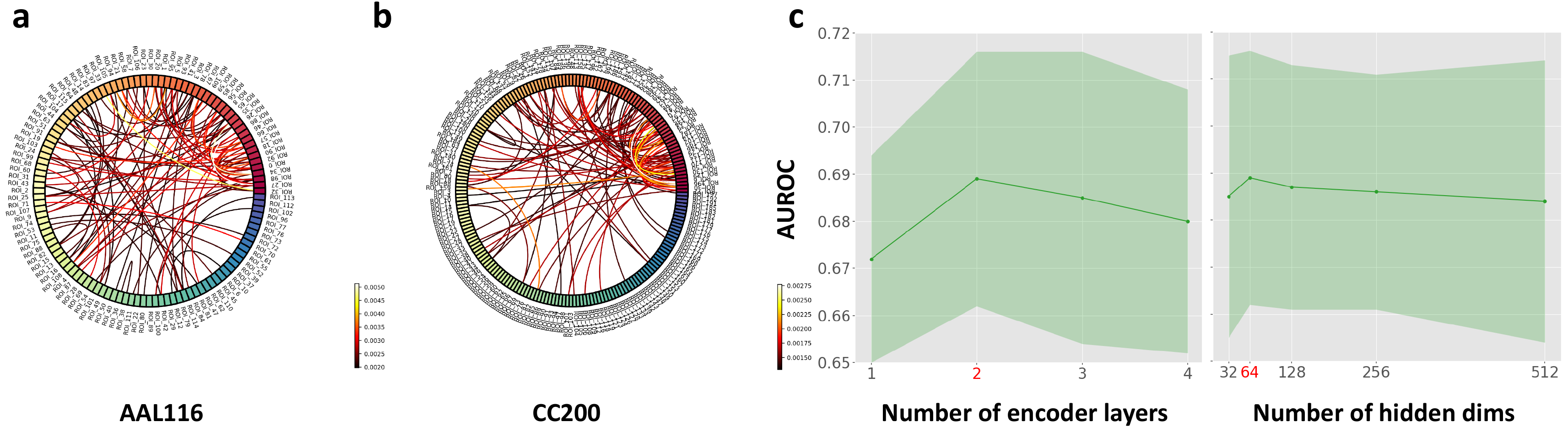}}
\caption{Discriminative functional connections between brain regions under (a) AAL116 and (b) CC200 for ADHD diagnosis. (c) Sensitivity analysis of AUROC with respect to the number of encoder layers and the hidden dimension $d$. }
\label{fig:roi}
\end{figure}

Figure~\ref{fig:roi}(a) and (b) further visualize the discriminative brain region connections identified by MADCLE under the AAL116 and CC200 parcellations, respectively, providing a qualitative visualization of connectivity patterns emphasized by MADCLE under different parcellations.
Figure~\ref{fig:roi}(c) presents a sensitivity analysis of MADCLE with respect to the number of Transformer encoder layers and the hidden dimensionality. The AUC remains relatively stable across a broad range of settings, indicating that the proposed framework is not overly sensitive to architectural hyperparameters. Performance peaks at a moderate network depth, while deeper architectures do not yield additional gains, suggesting that excessive model complexity is unnecessary for capturing disease-relevant FC patterns. Overall, these results demonstrate that MADCLE is robust to reasonable variations in model configuration.
\section{Conclusion}
In this paper, we proposed MADCLE, a disentangled multi-atlas FC learning framework for brain disorder identification. MADCLE learns atlas-wise disease-related embeddings and encourages their cross-atlas consistency, while separately modeling covariate-related and atlas-dependent factors through covariate supervision, reconstruction, and decorrelation constraints. Experiments on ADHD-200 and ADNI suggest that this structured representation learning strategy can achieve competitive or improved performance over single-atlas and multi-atlas baselines under heterogeneous parcellation schemes.
These results support the value of jointly modeling cross-atlas consistency and atlas-dependent residual variation. However, the current formulation encourages consistency between atlas-wise disease embeddings rather than learning a single explicit shared latent variable, and the biological meaning of atlas-dependent factors remains to be further validated. Future work will explore stronger shared latent modeling, broader validation across atlases and cohorts, and more explicit neurobiological interpretation.


%
%
%
\newpage
\bibliographystyle{splncs04}
\bibliography{reference}
\end{document}